\documentstyle[11pt,revpasp,psfig]{article}
\def\edcomment#1{\iffalse\marginpar{\raggedright\sl#1\/}\else\relax\fi}
\reversemarginpar

\begin{document}
\title{The Massive Stellar Content in Starburst  Galaxies and its Impact on Galaxy Evolution}
\author{Rosa M. Gonz\'alez Delgado}
\affil{Instituto de Astrof\'\i sica  de Andaluc\'\i a (CSIC), Apdo. 3004, 18080 Granada, Spain. rosa@iaa.csic.es}

\begin{abstract}

Starburst galaxies are powered by massive stars. These stars dominate the heating and enrichment with heavy elements of the interstellar medium, gas out of which new stars form. Thus, high-mass stars, and in consequence starburst galaxies, are an important (in some cases the dominant) energy source for the evolution of galaxies and the universe. In this contribution I review techniques to derive the massive stellar content in starburst galaxies and their evolutionary state. They are based on the analysis of their integrated light with evolutionary synthesis models. The massive stellar population is derived in a self-consistent way using the continuum and {\it stellar} wind resonance lines in the ultraviolet, the photospheric H Balmer and HeI lines and the {\it nebular} emission lines  at optical wavelengths. Comparison with observations provides constraints on the most recent star formation history in starburst galaxies and their evolutionary state.    

\end{abstract}

\section{Starburst Galaxies and its Cosmological relevance}

Massive stars are defined as having masses higher than 10 M$\odot$ when they are at the zero age main sequence (ZAMS). They are very luminous and dominate the heating of the interstellar medium (ISM), gas out of which new stars form. They emit photons that can ionize the surrounding interstellar gas and deposit mechanical energy both via stellar winds and supernovae. Most of the heavy elements are formed in massive stars, and these metals are dispersed in the ISM when these stars finally explode as supernova. Therefore, massive stars are an important source (sometimes the dominant one) for the evolution of a galaxy.
Massive stars are formed in the arms of normal spiral galaxies. However, the star formation rate is low ($\simeq$ 1 M$\odot$ yr$^{-1}$) and the gas available in these galaxies is enough to sustain the rate for many Gyr. On the contrary, much higher rate of star formation occurs in the galaxies know as starburst galaxies. 

A starburst is defined as a brief episode of intense star formation that is taking place in small regions (100 to 1000 pc) of the galaxy and dominates the overall luminosity of the galaxy (Weedman 1983; Heckman 1998). The star formation rate is so high (10-100 M$\odot$ yr$^{-1}$ and in some cases up to 1000 M$\odot$ yr$^{-1}$) that galaxies can be in this phase only for a small fraction of the age of the universe ($\leq$10$^8$ yr) because the gas available in these galaxies cannot sustain this rate for a longer time. This definition covers galaxies with a very wide
 range of properties, from blue compact dwarfs, nuclear starbursts  to ultraluminous IRAS starbursts.  Even giant HII regions as 30 Doradus can be considered as a mini starburst. Typical masses 
(bolometric luminosities) range from 10$^6$ to 10$^{10}$ M$\odot$ (10$^9$ to 10$^{14}$ L$\odot$), 
corresponding the lowest limit to the mass of super-star-clusters and the highest limit to the 
mass of the infrared-luminous galaxies (Leitherer 1996). 
Starburst galaxies are ideal laboratories in which to explore fundamental questions about the local and 
global processes and effects of star formation. For example, is the initial mass function the
same in starburst galaxies as in quiescent star formation in the Milky Way? Is there a connection 
between starburst activity and active galactic nuclei? What can starbursts teach us about the process
by which the first galaxies formed?  

Starbursts are a significant energy source of our local universe. The four most luminous circumnuclear starburst (M 82, NGC 253, M 83 and NGC 4395) can account about 25 $\%$ of the high mass star formation rate in a volume of 10 Mpc radius from us (Heckman 1997).  Starbursts play also an important role in the understanding of the formation and evolution of galaxies when the universe was very young.  In the last three years, a significant population of normal star-forming galaxies have been discovered at redshift higher than 2. Most of these galaxies were discovered with the Lyman-break technique (Steidel et al 1996; Lowenthal et al 1997), and a few of them serendipitously (e.g. Yee et al 1996; Ebbels et al 1996).  They are considered to be precursors of typical  present-day galaxies in an early actively-star-forming phase. One of the best case studied until now is the galaxy MS 1512-cB58 (Yee et al 1996), a galaxy at z=2.723 which is gravitational lensed. Its ultraviolet  (UV) spectrum shows high ionization absorption features (CIV $\lambda$1540 and SiIV $\lambda$1400) formed in the stellar winds of massive stars (Pettini et al 1997); these characteristics are very similar to local starbursts  so that it can be classified as a starburst galaxy. Due to this similarity, local starburst galaxies can be considered as benchmark objects to understand the massive star formation in the early stage of the formation of the universe.  

Therefore, to study in detail local starbursts can help to understand how the star formation proceeded in the early stages of the universe. In this paper, I review three techniques that can be used to estimate the evolutionary state of the starbursts and their massive stellar content. They are based on the spectral morphology of the integrated light of a starburst at ultraviolet and optical wavelengths, and on the concept of the evolutionary synthesis technique.

\section{The Concept of Evolutionary Synthesis Technique.}

Nearby starbursts are at distances that individual stars are not detected. One of the closer starburst is in the super star clusters in the irregular galaxy NGC 1569 (Arp \& Sandage 1985). The typical size of these super star clusters is about 1 pc (O'Connell, Gallagher \& Hunter 1994), that a distance of 2.5 Mpc  has an angular size less than 0.1 arcsec. Therefore, the stellar content has to be determined through the analysis of the integrated light of the whole starburst.
  
One powerful tool for interpreting the integrated light of  starbursts is the evolutionary synthesis, that was introduced by Tinsley (1968). This  technique makes a prediction for the synthetic spectrum of a
stellar population taking as a free  parameter the star formation history of the
starburst (age, initial mass function, star formation rate, etc). The basic ingredients are: (i) stellar evolutionary tracks; (ii) spectral libraries, either empirical spectra or theoretical model atmospheres. The way that the technique works is this: (i) initally the population of stars is distributed along the ZAMS according to an initial mass function (IMF); (ii) the evolution of the stars from the ZAMS until they become insignificant in their energy production is followed; (iii) the spectrum of each star is computed; (iv) the resulting spectrum is obtained adding for each epoc the contribution of each of the stars. Thus, the spectrum and other related observables (colors, luminosities, number of ionizing photons, supernova rate, etc) are predicted. If there is agreement between the observations and the predicted physical properties, 
constraints on the star formation history (in timescales between 1 Myr to several Gyr) can be determined (see e.g. Mas-Hesse \& Kunth 1991; Bruzual \& Charlot 1993; Leitherer \& Heckman 1995). However, the solution obtained with this technique may not be unique, and only {\em consistency}
between models and observations can be derived. The power of the technique is very dependent on the reliability of the stellar models used as ingredients. 

One important question related with this technique is whether  evolutionary
synthesis models, and  the stellar evolution models they incorporate, are actually up to the
challenge of interpreting multi-wavelength data. To answer this, I will present a  multi-wavelength study of the starburst galaxy NGC 7714. 

\section{The Spectral Morphology of a Starburst.}

Starburst galaxies are characterized by showing a nebular emission line spectrum at optical and an absorption line spectrum at ultraviolet wavelengths. 
This is called "the spectral dichotomy picture of a starburst galaxy" 
(Leitherer 1997). This picture reflects that starbursts are powered by massive 
stars. These stars emit photons with energies of a few 
tens of eV which are absorbed and re-emitted in their stellar winds, producing 
ultraviolet resonance transitions, and then, an ultraviolet absorption line spectrum (Figure 1a).
However, most of the ionizing photons emitted by the stars can travel tens of pc 
from them and be absorbed by the surrounding interstellar gas. Then, this ionized gas cools down
emitting a nebular emission line spectrum, mainly at optical wavelengths (Figure 1b). 

However, at the Balmer jump the spectra can show absorption features formed in the photosphere 
of massive stars. O, B and A stars can dominate the optical continuum emission of the starburst galaxies.
The spectra of the early-type stars is characterized by strong hydrogen Balmer and 
neutral helium absorption lines with only very weak metallic lines (Walborn \&
Fitzpatrick 1990). However, the detection of these stellar features at optical wavelengths
in the spectra of starburst galaxies is a difficult business because H and HeI absorption
features are coincident with the nebular emission lines that mask the absorption features.
Even so, the higher order terms of the Balmer series and some of the HeI lines are
detected in absorption in many starburst galaxies (Storchi-Bergmann, Kinney \& Challis 1995; 
Gonz\'alez-Delgado et al 1995), in Seyfert 2 galaxies with their ultraviolet and optical 
continuum dominated by a nuclear starburst (Shields \& Filippenko 1990; Gonz\'alez Delgado et al 1998b;
Cid Fernandes, Storchi-Bergmann \& Schmitt 1998) or even in the spectra of giant 
HII regions (e.g. NGC 604, Terlevich et al 1996).

In the following sections I will show how to obtain information about the massive stellar content of starburst galaxies and their evolutionary state, based on the spectral morphology of these objects and using evolutionary synthesis models.

\begin{figure}

\psfig{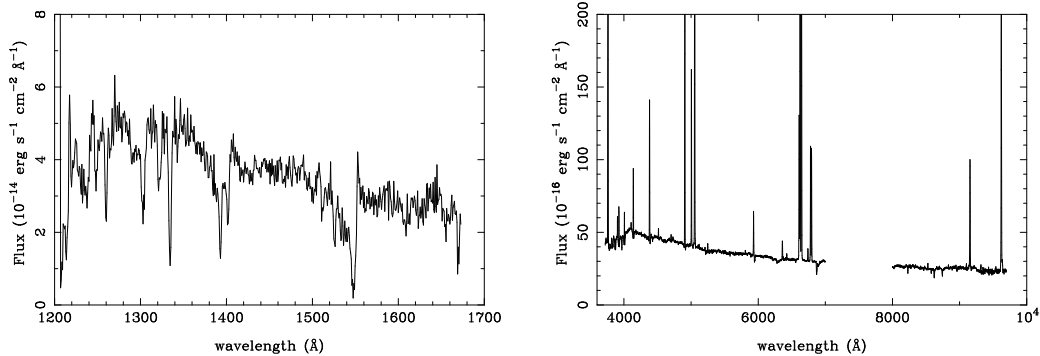}
\caption[fig]{(a) HST+GHRS spectrum of NGC 7714  (b) Ground-based optical spectrum of NGC 7714. The spectrum shows the so-called called "spectral dichotomy picture of a starburst galaxy": a nebular emission line spectrum at optical and an absorption line spectrum at ultraviolet wavelengths. }
 \end{figure}

\section{The Optical Emission Line Spectrum of a Starburst.}

The emission line spectrum of a starburst depends on the radiation field from the ionizing stellar cluster, the electron density and chemical composition of the gas. These parameters are taken as input to a photoionization code, that resolves the
ionization-recombination and heating-cooling balances, and predicts the ionization structure of the nebula, the electron density and the intensity of the emission lines.  The evolutionary state of the starburst and the massive stellar content can be derived using the spetral energy distribution generated by a stellar evolutionary synthesis code as the radiation field. This technique has been applied very successfully to constrain the star formation history and the evolutionary state of extragalactic HII regions, starbursts (Garc\'\i a-Vargas, Bressan, \& D\'\i az 1995; Stasi\'nska  \& Leitherer 1996; Garc\'\i a-Vargas et al 1997; Gonz\'alez Delgado et al 1999) and active nuclei of galaxies (Terlevich \& Melnick 1985; Cid-Fernandes et al 1992).

I show how this technique works by applying it to the prototypical nuclear starburst NGC 7714 (Weedman et al 1981). The spectral energy distribution is generated by the evolutionary synthesis
code  developed by Leitherer and collaborators (Leitherer et al 1999). The code uses the most recent stellar evolutionary models of the Geneva group (Schaller et al 1992; Maeder \& Meynet 1994), and the stellar atmospheres grid compiled by Lejeune et al (1996). The spectral energy distribution was generated using the 0.4 Z$\odot$ metallicity
tracks because the chemical abundance derived from the emission lines in NGC
7714 is close to half solar (Gonz\'alez-Delgado et al 1995), and assuming two different star formation
scenarios (instantaneous burst and continuous star formation). In both cases, a
Salpeter IMF with an upper and lower mass limit cut-off of 80 M$\odot$ and 1
M$\odot$, respectively, was assumed. 

The spectral energy distribution is taken as input to the photoionization code
CLOUDY (version 90.04, Ferland 1997). It is assumed that the nebular gas is ionization bounded and
spherically distributed around the ionizing cluster with a constant density.  The inner radius of the nebula is taken at 3 pc, but the outer radius is determined by the
ionization front. The chemical composition of the gas is fixed to the values calculated from the optical emission lines, and the ionizing photon luminosity is fixed to log(Q)=52.9 ph s$^{-1}$,
as derived from the Balmer recombination lines. Models are computed taking the
filling factor as a free parameter. The change
in the filling factor is equivalent to changing the ionization parameter U,
defined as Q/(4$\pi$R$_s$N$_e$c); where Q is the ionizing photon luminosity,
N$_e$ the electron density, c the speed of light and R$_s$ the Str\"omgren
radius. The average U is proportional to ($\phi^2$ N$_e$ Q)$^{1/3}$, where $\phi$
is the filling factor. 
First, the ionization parameter is derived using the ratio
[SII]$\lambda$6716+6731/H$\beta$. This ratio is a good calibrator of U for
continuous star formation and for burst models (Figure 2a). The observed ratio indicates an ionization parameter of -2.9,
corresponding to a filling factor of 0.001. Then, the other emission lines
can constrain the star formation scenario (burst or continuous star formation) and
the age of the starburst (Figure 2b). Thus, photoionization models predict that the nuclear starburst in NGC 7714 was formed instantaneously  4.5 Myr ago. 

\begin{figure}

\psfig{file=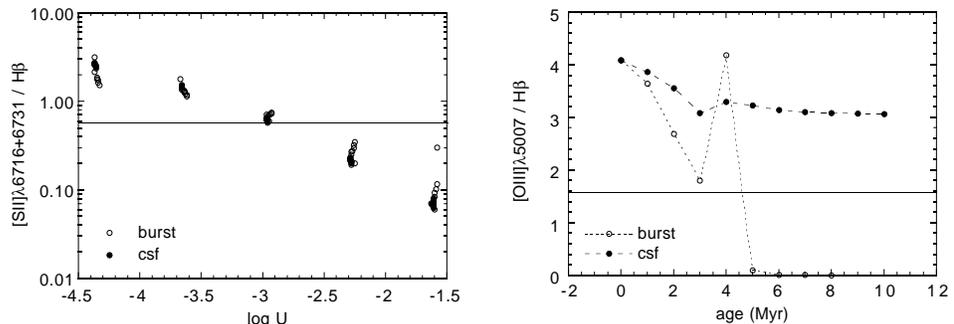,height=4.5cm}

\caption[fig]{Predicted emission line ratio: (a)
[SII]$\lambda$6716+6732/H$\beta$ as a function of the average ionization
parameter generated with CLOUDY, taking as input the spectral energy
distribution from the evolution of a burst (open symbols) and
continuous star forming regimes (full symbols);  (b) [OIII]$\lambda$5007/H$\beta$ as a function of age, assuming a filling factor of 0.001. The observed values are indicated by a horizontal line. 
These ratios are compatible with a burst 4.5 Myr old, and they  exclude the continuous star formation models.}
 
\end{figure}

\section{The Ultraviolet Absorption Line Spectrum of a Starburst.}

A second independent method that can be used to constrain the star formation history and the evolutionary state of a starburst is based on the profile of the ultraviolet wind resonance lines. Massive hot stars develop strong stellar winds due to radiation pressure in
ultraviolet resonance lines (Morton 1967). Typical wind
velocities in O stars are about 2000 km s$^{-1}$ to 3000 km s$^{-1}$
(Groenewegen, Lamers \& Pauldrach 1989). As a result, all
strong ultraviolet lines in the spectra of O stars originate predominantly in
the outflow, and have blueshifted absorptions. The profile shapes reflect the
stellar mass-loss rates, which are a strong function of the stellar luminosity
(Castor, Abbott \& Klein 1975). Since there exists a well-defined stellar
mass-luminosity relation, the line profiles ultimately depend on the stellar
mass, and,  for a stellar population, on the IMF and SF history. 

The strongest stellar wind resonance features are OVI $\lambda$1034, NV
$\lambda$1240, SiIV $\lambda$1400 and  CIV $\lambda$1550. In massive stars, these lines form above in the stellar wind, as a blueshifted absorption in stars
with weak winds, or as a P Cygni profile if the wind density is sufficiently
high. OVI and CIV are strong lines in O stars of all luminosity classes
(Walborn, Bohlin \& Panek 1985; Gonz\'alez Delgado, Leitherer \& Heckman
1997a). In contrast, SiIV is luminosity dependent, and only  blue supergiant
stars produce a strong P Cygni profile (Walborn et al 1985). The recombination  line HeII $\lambda$1640 can also be formed in very massive O and WR stars with 
very dense winds. Evolutionary synthesis models show that the profile of these lines depend 
on the IMF parameters and the evolutionary state of the starburst (Leitherer, Robert \& Heckman 1995; Gonz\'alez Delgado et al 1997a). This
technique has been succesfully applied to starburst galaxies (Conti, Leitherer \& Vacca 1996; 
Leitherer et al 1996; Gonz\'alez Delgado et al 1998a; Gonz\'alez Delgado et al 1999) and Seyfert
galaxies (Heckman et al 1997; Gonz\'alez Delgado et al 1998b). In all these
cases, the method has constrained the age and the mass spectrum of the young
population. In next section I apply this technique to the nuclear starburst of NGC 7714. 

\subsection{The nuclear starburst in NGC 7714.}

The nucleus of NGC 7714 was observed with HST+GHRS with square aperture of 1.74$\times$1.74 arcsec$^2$ and with a dispersion of 0.57 \AA/pix. Figure 3a  shows the nuclear morphology of the central 2 arcsec of the starburst.  The spectrum shows resonance lines formed in the interestellar medium of the galaxy (e.g. SiII $\lambda$1260, OI $\lambda$1302, CII $\lambda$1335, SiII $\lambda$1527, FeII $\lambda$1608 and AlII $\lambda$1607) and in the winds of massive stars (NV $\lambda$1240, SiIV $\lambda$1400, CIV $\lambda$1540 and HeII $\lambda$1640).  Fitting the profile of the SiIV and CIV lines with the evolutionary synthesis models, it is found that the profiles of these lines are compatible with an instantaneous burst  5 Myr old (Figure 3b).  Thus, this technique can constrain the evolutionary state of the starburst giving the same result as the fit of the emission line spectrum. 

The ultraviolet continuum luminosity can also be used to estimate the bolometric luminosity and the mass of the starburst, but first an estimation of the extinction is required. Evolutionary synthesis models also show that the ultraviolet
energy distribution arising from a starburst has a spectral index, $\alpha$, in
the range -2.6 to -2.2 (F$_\lambda\propto$ $\lambda^{\beta}$), if the burst is
less than 10 Myr old (Leitherer \& Heckman 1995). This spectral index is rather independent of the  metallicity and IMF. Therefore, any deviation from the predicted spectral index can be
attributed to reddening.  The spectral index derived from the UV spectrum of NGC 7714 indicates that the internal extinction in the nuclear starburst of NGC 7714 is low (E(B-V)=0.03) and the luminosity derived at 1500 \AA\ is 10$^{39.9}$ erg s$^{-1}$ \AA$^{-1}$. This luminosity is produced by a starburst of 5-10$\times$10$^6$ M$\odot$ and bolometric luminosity  5-10$\times$10$^9$ L$\odot$. The HeII $\lambda$1640 luminosity indicates a stellar population of $\sim$2000 Wolf-Rayet stars. Models predict a nuclear supernova rate of 0.001 yr$^{-1}$, which is two orders of magnitud lower than the value estimated previously based in the X ray emission of the whole galaxy (Weedman et al 1981). That rate is about the same as the disk-integrated rate observed in normal spirals. However, what distinguishes NGC 7714 and other nuclear starbursts is the concentration of supernova events in the nucleus.

We conclude that evolutionary synthesis models are able to interpret multi-wavelength data of starbursts because the two techniques applied here to the nucleus of NGC 7714 give the same solution.

 \begin{figure}

\psfig{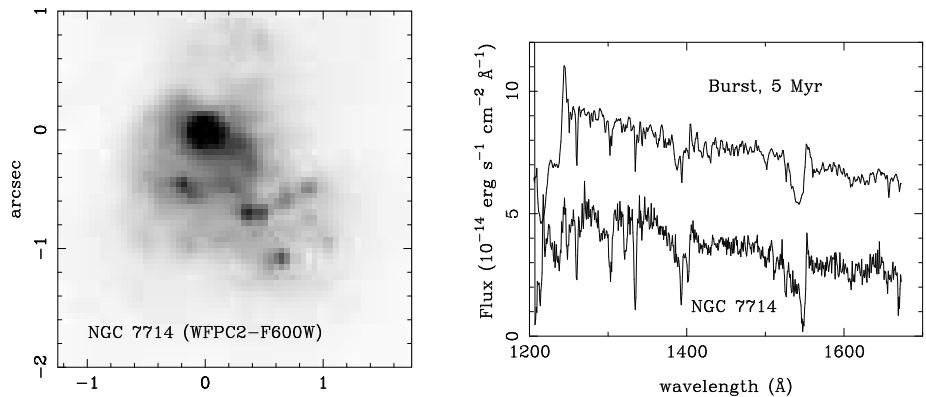}
\caption[fig]{(a) HST+WFPC2 (F600W) image  of the nuclear starburst of NGC 7714. The nucleus is clearly extended, and resolved into severals knots.  (b) Dereddened HST+GHRS spectrum of NGC 7714, and the synthetic 5 Myr burst model (in relative units). The IMF slope is
Salpeter and $\rm M_{up}=100~M\odot$.}
 
\end{figure}

\section{The Hydrogen Balmer and HeI absorption lines.} 

Signatures of massive stars are also detected in the near-ultraviolet spectra of starbursts (Figure 1b) and
in the nuclear spectra of Seyfert 2 galaxies (Figure 4).  
These features can be seen in absorption 
because the strength of the Balmer serie in emission decreases rapidly with decreasing 
wavelength, whereas the equivalent width of the stellar absorption lines is constant 
or encreases with wavelength. Thus, the net effect is that H$\alpha$, H$\beta$ and H$\gamma$
are mainly seen in emission, but the higher order terms of the Balmer series are seen 
in absorption. Very often, H$\alpha$, H$\beta$, H$\gamma$ and H$\delta$ show absorption wings
over-imposed on the nebular emission and  a similar effect can occur for the HeI lines. The strongest 
nebular emission occurs in the HeI $\lambda$5876 and HeI $\lambda$6678 whereas the stellar 
absorption features are very weak. Meanwhile, the equivalent width of the nebular emission 
of HeI $\lambda$4471 and HeI $\lambda$4026 ($\leq$ 6 \AA,  Izotov, Thuan \& Lipovetsky 1997) is similar or even smaller than He stellar absorption lines; thus, they can be easily detected in absorption.

Therefore, evolutionary synthesis models that predict the strength of the photospheric  H Balmer and HeI lines can also be used to study the evolutionary state and star formation history of starburst galaxies. 

\begin{figure}

\psfig{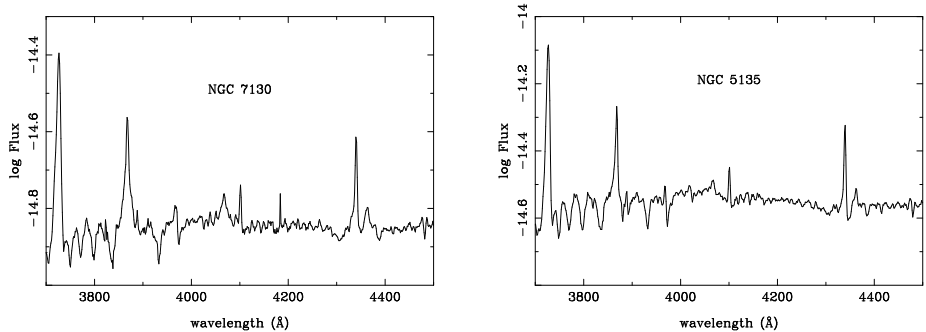}
\caption[fig]{Ground-based optical spectrum of the nucleus of the Seyfert 2 galaxies NGC 7130 (a) and NGC 5135 (b). The high-order series Balmer and HeI lines detected in absorption indicate the existence of a starburst in the nucleus of these galaxies. }

\end{figure}

\subsection{The stellar library.}  

It is built with synthetic spectra 
covering five spectral ranges between 3700 and 5000 \AA. The spectra include the profiles
 of the Balmer (H13, H12, H11, H10, H9, H8, H$\delta$, H$\gamma$ and H$\beta$) 
and HeI ($\lambda$3819, $\lambda$4009, $\lambda$4026, $\lambda$4120, 
$\lambda$4144, $\lambda$4387, $\lambda$4471, $\lambda$4922, and $\lambda$5876) 
lines with a sampling of 0.3 \AA. The grid spans a range of effective temperature
 50000 K $\geq$ T$_{eff}$ $\geq$ 4000 K, and  gravity 0.0 $\leq$ log g $\leq$ 5.0 at solar
 metallicity. For T$_{eff}$ $\geq$ 25000 K, NLTE stellar atmosphere models are computed 
using the code TLUSTY (Hubeny 1988). These models together with Kurucz (1993) LTE stellar 
atmosphere models (for T$_{eff}\leq$ 25000 K) are used as input to SYNSPEC (Hubeny, Lanz \& Jeffery 1995), the program that solves the radiative transfer equation. Finally, the synthetic spectrum is obtained after performing the rotational and instrumental convolution.

\subsection{The Evolutionary Synthesis Models of the H and HeI lines.}

This library is used as input to an evolutionary synthesis code (Starburst 99, Leitherer et al 1999).
The models are optimized for galaxies with active star formation and they predict the strength of the H Balmer and He I lines for a burst with an age ranging from 10$^6$ to 10$^9$ yr, and 
different assumptions about the IMF. Continuous star formation models lasting for 
100 Myr are also computed.  

\begin{figure}

\psfig{file=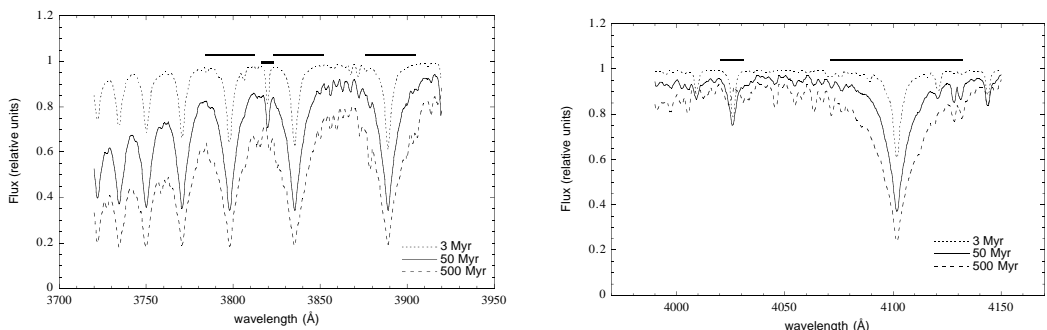,height=4.5cm}
\caption[fig]{Synthetic spectra from 3700 to 4200 \AA\ predicted for an instantaneous 
burst of 10$^6$ M$\odot$ formed following a Salpeter IMF between M$_{low}$= 1 M$\odot$ 
and M$_{up}$= 80 M$\odot$ at the age 3, 50 and 500 Myr.}
 
\end{figure}

It is found that the Balmer and HeI line profiles are sensitive to the age (Figure 5), except 
during the first four Myr of the evolution, when the equivalent widths of these lines 
are constant (Figure 6). At these early stages of the evolution, the lines profiles are 
also sensitive to the IMF (Figure 7). However, strong H Balmer and HeI lines are predicted even 
when the low mass cut-off of the IMF is as high as 10 M$\odot$. The equivalent width 
of the Balmer lines ranges from 2 to 16 \AA\ and the HeI lines from 0.2 to 1.2 \AA.  
During the nebular phase (cluster younger than  about 10 Myr), H$\beta$ ranges from 2 to 
5 \AA\ and HeI $\lambda$4471 between 0.5 and 1.2 \AA. The strength of the lines is 
maximum when the cluster is a few hundred (for the Balmer lines) and a few ten 
(for the HeI lines) Myr old. In the continuous star formation scenario, the strength 
of the lines increases monotonically with time; however, the lines are weaker than 
in the burst  models due to the dilution of the Balmer and HeI lines by the 
contribution from very massive stars. 

\begin{figure}

\psfig{file=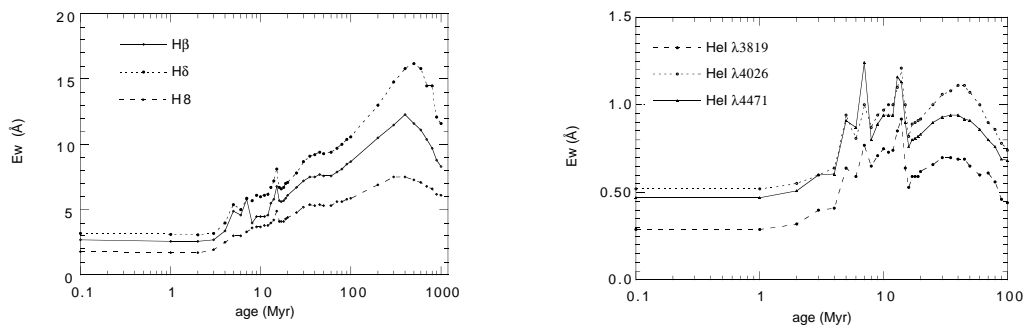,height=4.5cm}
\caption[fig]{Equivalent widths of the Balmer lines (a) and HeI lines (b) measured in the evolutionary synthesis models for an instantaneous burst formed following a Salpeter IMF between M$_{low}$= 1 M$\odot$ and M$_{up}$= 80 M$\odot$.}
 
\end{figure}

These models can be used to date starburst and post-starburst galaxies until 1 Gyr.  Due to the high spectral resolution of the profiles it is possible to reproduce the absorption wings observed in regions of recent star formation, to estimate the effect on the nebular emission lines by underlying 
absorption and the evolutionary state of the starburst.

\begin{figure}

\psfig{file=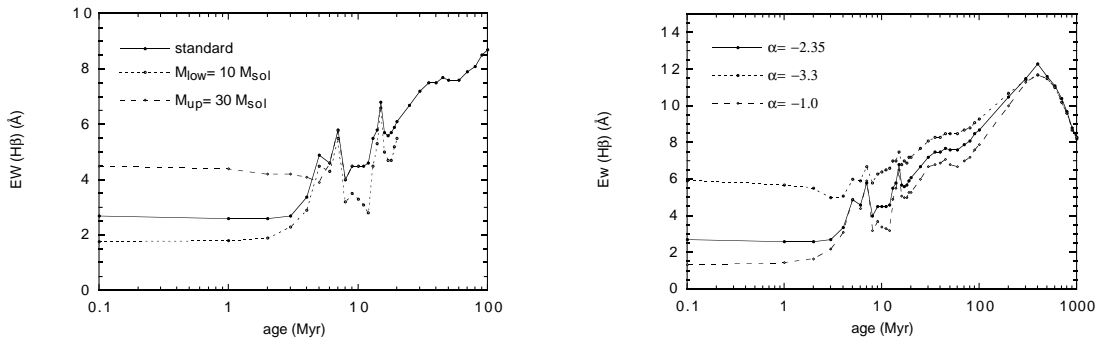,height=5.0cm}
\caption[fig]{Equivalent width of H$\beta$ for an instantaneous burst at solar metallicity formed following different assumptions about the IMF parameters.}
 
\end{figure}

\subsection{Dating Super Stars Clusters.}

In this section, I show how to date the evolutionary state of super star clusters (SSCs) using synthesis profiles of the H and HeI lines.
SSCs are characterized to be very luminous and compact objects. 
There are suggestions that indicate SSCs represent the present-day analogs of young 
globular clusters because they have masses and luminosities comparable to those of 
evolved globular clusters in the Milky Way (Ho \& Filippenko 1996; O'Connell, Gallagher
\& Hunter 1994). They represent the basic mode of star formation in starburst galaxies
and are the building blocks of these objects. The clusters can be aproximated by a coeval 
and single metallicity stellar population.

Two of the closest SSCs are in the dwarf irregular galaxy NGC 1569. HST images of NGC 1569, 
suggest that the SSC B has an age of 15-300 Myr (O'Connell et al 1994) and ground-based optical spectra suggest an age of 10 Myr (Gonz\'alez Delgado et al 1997b). The last result is based on the 
analysis of the optical spectral energy distribution and on the strength of the CaII triplet 
(see also Prada et al 1994). To test these models, I have compared the synthetic profiles
 of the Balmer lines with the observations. The Balmer lines are partially filled 
with nebular emission; therefore the fitting has to be done based on the wings of the 
absorption features. This effect is less important for the higher terms of the Balmer 
series because the 
nebular emission drops quickly with decreasing wavelength. Figure 8 plots the observed
 lines and the synthetic models for a burst of 10 and 50 Myr at Z$\odot$/4 metallicity 
(assuming Salpeter IMF, M$_{low}$=1 M$\odot$ and M$_{up}$=80 M$\odot$). The profiles 
indicate that the Balmer lines are more compatible with a burst of 10 Myr old than with
 the 50 Myr old. Ages older than 10 Myr produce profiles which are wider than the 
observed one. This comparison confirms the result of Gonz\'alez Delgado et al (1997b)
that the age of the SSC B is about 10 Myr and shows that this technique can discriminate 
well between a young and an intermediate age population, and to estimate the evolutionary 
state of SSCs.

\begin{figure}

\psfig{file=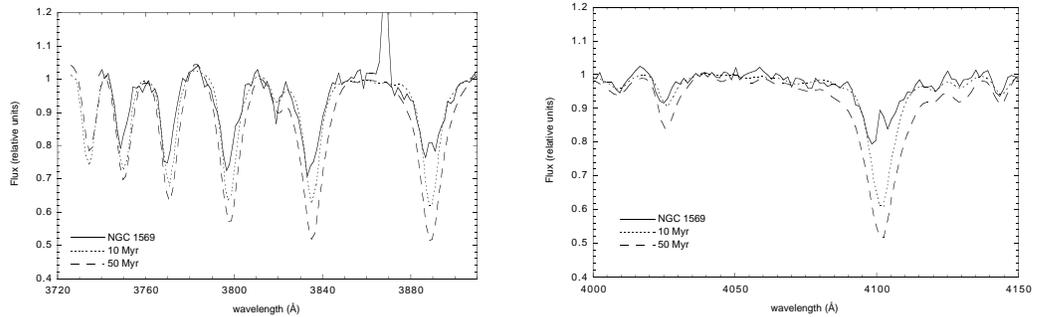,height=4.5cm}
\caption[fig]{Normalized optical spectrum of the SSC B in NGC 1569 (full line). The synthetic normalized spectra of an instantaneous burst 10 Myr (dotted line) and 50 Myr old (dashed line) formed following a Salpeter IMF at Z$\odot$/4 metallicity are ploted. Models are smoothed to the resolution of the observations.}
 
\end{figure}

\section{Summary}

A starbursts is a brief episode of intense star formation that is taking place in a small region of a galaxy and dominates its overall luminosity. Most of the radiative properties of the starburst galaxies are determined by their massive stellar content. High-mass stars are responsible for the spectral morphology of a starburst that shows a nebular emission line spectrum at optical and an absorption line spectrum at ultraviolet wavelengths. Nearby starburst galaxies are considered as local analogs of active star forming galaxies at high redshift. Thus, nearby starbursts are very good laboratories to explore spectroscopic techniques than can be later to apply in high redshift galaxies to derive their massive stellar content and their evolutionary state.  I have reviewed several techniques based on evolutionary synthesis models that predict the stellar wind resonance lines at ultraviolet, the photospheric H Balmer and HeI lines and the nebular emission lines at optical wavelengths. The three techniques are able to constrain the star formation history in nearby starbursts and to predict their content in massive stars and their evolutionary state.   

\section{Acknowledgement}

I would like to thank the organizers of this conference, in particular John Beckman and M\'onica Murphy, for their financial support and for making this meeting enjoyable, and to my collaborators Marisa Garc\'\i a Vargas and Claus Leitherer for their important contributions to the realization of several parts of this work. This work was supported by the Spanish DGICYT grant PB93-0139.

\end{document}